\documentclass[12pt]{article}
\usepackage{geometry} 
\usepackage{graphicx} 
\usepackage{amsmath} 
\usepackage{amsfonts} %
\usepackage{cite}          
\usepackage{hyperref}  

\usepackage{tikz}
\usepackage{pgfplots}
\usetikzlibrary{decorations.markings}
\usetikzlibrary{patterns}
\def\cc{\mbox{c}_{\scalebox{0.36}{\bf CFT}}}

\title{Analysis for Lorentzian conformal field theories through sine-square deformation}
\author{Xun LIU$^\ast{}^\dag$ and Tsukasa TADA${}^\dag{}^\ddag$}
\date{
}

\begin{document}
\maketitle
\centerline{\it ${}^\ast$Department of Physics, The University of Tokyo} 
\centerline{\it 7-3-1 Hongo, Bunkyo-ku, Tokyo 113-0033, Japan}

\centerline{\it ${}^\dag$RIKEN Nishina Center for Accelerator-based Science} 
\centerline{\it Wako, Saitama 351-0198, Japan}

\centerline{\it ${}^\ddag$RIKEN Interdisciplinary Theoretical and Mathematical Sciences (iTHEMS)}
\centerline{\it Wako, Saitama 351-0198, Japan}
\renewcommand{\L}{{\cal L}}
\renewcommand{\l}{\mathfrak l}
\begin{abstract}
We reexamine two-dimensional Lorentzian conformal field theory using the formalism previously developed in a study of sine-square deformation of Euclidean conformal field theory. We construct three types of Virasoro algebra. One of them reproduces the result by L\"uscher and Mack, while another type exhibits the divergence in the central charge term. The other leads the continuous spectrum and contains no closed time-like curve in the system.
\end{abstract}

In \cite{Ishibashi:2015jba,Ishibashi:2016bey}, it was shown that the sine-square deformation (SSD) \cite{Gendiar:2008udd} of conformal field theory \cite{Katsura:2011ss} can be understood in terms of the new quantization scheme called the ``dipolar quantization.'' So far, the analysis has been limited in the Euclidean theories. In this note, we would like to extend our analysis to Lorentzian conformal field theories \cite{Luscher:1976,Mack:1988nf,Luscher:1974ez}
\footnote{Some of the earlier studies on SSD are \cite{1104.1721,1108.2973}. In the context of string theory and conformal field theory, it was studied in \cite{Tada:2014kza,Tada:2014wta}. More recent studies include \cite{Okunishi:2015dfa,Wen:2016inm,Okunishi:2016zat,Tamura:2017vbx,Tada:2017wul,Wen:2018vux,Wen:2018agb,Tada:2019rls}.}.

\noindent{\bf Minkowski covering space}\ \ \ 
To consider conformal field theories in Minkowski spacetime, one need to introduce the universal covering space of the original Minkowski spacetime \cite{Luscher:1974ez}. In two-dimensional (2d) case, the universal covering space can be conveniently depicted as the surface of the cylinder of the unit radius and infinite length with the coordinate:
\begin{equation}
( \sin\sigma, \cos\sigma, \tau),
\end{equation}
which can be translated into the 2d Cartesian coordinates:
\begin{equation}
x^{0}=\frac{\sin\tau}{\cos\tau+\cos\sigma} \ , \ x^{1}=\frac{\sin\sigma}{\cos\tau+\cos\sigma}.
\end{equation}
As a matter of fact, there exists a much simpler expression
\begin{equation}
x^{0}\pm x^{1} = \tan\frac{\tau\pm\sigma}{2}, \label{eqn:xpmtausigma}
\end{equation}
which prompts us to introduce a new set of coordinates:
\begin{equation}
x^{\pm}\equiv x^{0}\pm x^{1} , \quad  u^{\pm}\equiv \frac{\tau\pm\sigma}{2},
\end{equation}
thus yielding
\begin{equation}
x^{\pm}=\tan u^{\pm}.\label{eqn:xpmupm}
\end{equation}
The above relation maps the entire Minkowski spacetime to the Penrose diamond; the Minkowski spacetime is mapped to a diamond shape on the cylinder, which is the universal covering of the (Penrose) diamond.

\noindent{\bf conformal symmetry}\ \ \ The generators of the (2d) conformal transformation
\begin{eqnarray*}
& \hbox{[translation]}\ \ \ {\hat P}_{\mu}&=\frac{\partial}{\partial x^{\mu}}\ ,\\
& \hbox{[rotation]} \  \quad {\hat M}_{\mu\nu}&=g_{\mu\rho}x^{\rho}\frac{\partial}{\partial x^{\nu}}-g_{\nu\rho}x^{\rho}\frac{\partial}{\partial x^{\mu}} \ ,\\
& \hbox{[dilation]} \  \ \ \ \quad  {\hat D}&=x^{\mu}\frac{\partial}{\partial x^{\mu}} \ ,\\
 &\hbox{[SCT]}\ \  \ \qquad{\hat K}_{\mu} &=2g_{\mu\rho}x^{\rho}x^{\nu}\frac{\partial}{\partial x^{\nu}}-(x\cdot x)\frac{\partial}{\partial x^{\mu}}\ ,
\end{eqnarray*}
can be expressed explicitly in these coordinates as follows:
\begin{eqnarray}
 {\hat P}_{0}&=&\frac{\partial}{\partial x^{+}}+\frac{\partial}{\partial x^{-}}=\cos^{2}u^{+}\frac{\partial}{\partial u^{+}}+\cos^{2}u^{-}\frac{\partial}{\partial u^{-}} \ , \label{eqn:defP0}\\
  {\hat P}_{1}&=&\frac{\partial}{\partial x^{+}}-\frac{\partial}{\partial x^{-}}=\cos^{2}u^{+}\frac{\partial}{\partial u^{+}}-\cos^{2}u^{-}\frac{\partial}{\partial u^{-}} \ , \label{eqn:defP1}\\
 {\hat D}&=&x^{+}\frac{\partial}{\partial x^{+}}+x^{-}\frac{\partial}{\partial x^{-}}=\frac12\sin 2u^{+}\frac{\partial}{\partial u^{+}}+\frac12\sin 2u^{-}\frac{\partial}{\partial u^{-}} \ , \label{eqn:defD}\\  
  {\hat M_{01}}&=&x^{+}\frac{\partial}{\partial x^{+}}-x^{-}\frac{\partial}{\partial x^{-}}=\frac12\sin 2u^{+}\frac{\partial}{\partial u^{+}}-\frac12\sin 2u^{-}\frac{\partial}{\partial u^{-}} \ , \label{eqn:defM01}\\  
 {\hat K}_{0}&=&(x^{+})^{2}\frac{\partial}{\partial x^{+}}+(x^{-})^{2}\frac{\partial}{\partial x^{-}}=\sin^{2}u^{+}\frac{\partial}{\partial u^{+}}+\sin^{2}u^{-}\frac{\partial}{\partial u^{-}} \ , \label{eqn:defK0}\\
    {\hat K}_{1}&=&-(x^{+})^{2}\frac{\partial}{\partial x^{+}}+(x^{-})^{2}\frac{\partial}{\partial x^{-}}=-\sin^{2}u^{+}\frac{\partial}{\partial u^{+}}+\sin^{2}u^{-}\frac{\partial}{\partial u^{-}} \ , \label{eqn:defK1}
\end{eqnarray}
where we set the metric as
\begin{equation}
g_{00}=1 \ , g_{11}=-1 \ , g_{10}=g_{01}=0 .
\end{equation}
Note that each generator is composed of ``chiral'' part that includes exclusively $x^{+}$ or $u^{+}$,  and ``anti-chiral'' part with only $x^{-}$ or $u^{-}$.

The (rescaled) Casimir invariant of the conformal transformation is
\begin{eqnarray}
C_{(2)} &\equiv& (\hat D)^{2} -\frac{1}{2} \left( \hat P_{0} \hat K_{0}+\hat K_{0} \hat P_{0} \right) +\left(\hat M_{01}\right)^{2} 
+ \frac{1}{2} \left( \hat P_{1}\hat K_{1}+\hat K_{1} \hat P_{1}\right) ;\label{eqn:C2def}
\end{eqnarray}
the adjoint action of the above generators on their linear combination,
\begin{equation}
X\equiv d\hat D+p^{0} \hat P_{0}+k^{0}\hat K_{0}+m^{01}\hat M_{01}+p^{1}\hat P_{1}+\hat k^{1}K_{1},
\end{equation}
 leaves the following combination of the coefficients invariant:
\begin{equation}
c^{(2)} =(d)^{2}-4p^{0}k^{0}+(m^{01})^{2}+4p^{1}k^{1}.\label{eqn:c2def}
\end{equation}
Taking advantage of the invariance of $c^{(2)} $ under the adjoint action, one can, in turn, classify the (adjoint) generators by the value of $c^{(2)} $.

 The purpose of this note is to construct a set of Virasoro algebras, $\cal L_{\kappa}^{(+)}$ and $\cal  L_{\kappa}^{(-)}$, in the way that
\begin{equation}
{\cal L}_{0}^{(+)}+{\cal  L}_{0}^{(-)},
\end{equation}
evokes a time translation of our choosing. We shall choose, as the time translation, the 2d conformal transformation, which can be classified by $c^{(2)} $ and each classification may be conveniently represented by, for example, the following generators:
\begin{eqnarray}
\hat M_{01}
\quad& &\quad[c^{(2)} =1] , \label{eqn:c21}\\
\hat P_{0}\quad & & \quad[c^{(2)} =0], \label{eqn:c20}\\
\frac{\hat P_{0}+\hat K_{0}}{2}  & &\quad[c^{(2)} =-1] . \label{eqn:c2-1}
\end{eqnarray}
Eq. (\ref{eqn:c21}) is sometimes called Rindler Hamiltonian while Eq.(\ref{eqn:c20}) would be the most naive Hamiltonian. Eq. (\ref{eqn:c2-1}) is known as L\"uscher-Mack Hamiltonian
\cite{Luscher:1974ez}:
 \begin{equation}
\frac{\hat P_{0}+\hat K_{0}}{2}=\frac{1+(x^{+})^{2}}{2}\frac{\partial}{\partial x^{+}}+\frac{1+(x^{-})^{2}}{2}\frac{\partial}{\partial x^{-}}=\frac{\partial}{\partial u^{+}}+\frac{\partial}{\partial u^{-}}=\frac{\partial}{\partial \tau}.
\end{equation}
For the sake of conciseness, we only present the analysis for the chiral part in the following; the same analysis trivially applies to the anti-chiral case.

As the first step to the construction of Virasoro algebra, we start with the corresponding Witt algebra constructed on the space of the differential operators, following the treatment in \cite{Ishibashi:2015jba, Ishibashi:2016bey}. We introduce the following set of differential operators:
\begin{equation}
\ell_{\kappa}^{(+)} \equiv -g(x^{+})f_{\kappa}(x^{+})\frac{\partial}{\partial x^{+}}, \label{eqn:ellkappadef}
\end{equation}
where $g$ is the function that corresponds to the choice of our time-translation. 
$f_{\kappa}$'s follow the determining equations
\begin{equation}
\ell_{0}^{(+)}f_{\kappa}(x^{+})=-\kappa f_{\kappa}(x^{+}), \label{eqn:eigeneqf}
\end{equation}
while we  can set $f_{0}(x^{(+)})$ to be unity,
\begin{equation}
\ell_{0}^{(+)}=-g(x^{+})\frac{\partial}{\partial x^{+}},
\end{equation}
without loss of generality.
Therefore, $f_{\kappa}$'s are not only defining components of $\ell_{\kappa}^{(+)}$ but also span the space where $\ell_{0}^{(+)}$ acts on, as the eigenfunctions. Eq. (\ref{eqn:eigeneqf}) can be readily  solved and we obtain the explicit expression for $f_{\kappa}(x^{+})$ as
\begin{equation}
f_{\kappa}(x^{+})=\exp\left(\kappa \int^{x^{+}} \frac{dx'}{g(x')} \right). 
\end{equation}

Further calculation requires specifying the function $g$, which is to choose the time translation.
Should we choose any of the conformal transformations (\ref{eqn:defP0})-(\ref{eqn:defK1}) as the time translation, it is suffice to consider a quadratic function for $g$:  
\begin{equation}
    g(x)=ax^2+bx+ c. \label{eqn:gdef}
\end{equation}
Then, we obtain the expression of $f_{\kappa}$ , assuming $a\neq0$, as 
\begin{equation}
    f_{\kappa}(x)=\exp{(\frac{\kappa}{a(x_{\langle+\rangle}-x_{\langle -\rangle})}\ln{\frac{x-x_{\langle+\rangle}}{x-x_{\langle -\rangle}}})}
\end{equation}
in terms of the two root of the quadratic function $g$:
\begin{equation}
    x_{\langle\pm\rangle}=\frac{-b \pm \sqrt{\Delta}}{2a},
\end{equation}
where $\Delta$ is the discriminant $b^2-4ac$ \footnote{From the expression of $c^{(2)}$ (\ref{eqn:c2def}), one can see the value of the discriminant $\Delta$ coincide with $c^{(2)}$ for certain conformal transformations.}.

If $\Delta=0$, the quadratic function $g$ becomes degenerate: $g(x)=a(x+\frac{b}{2a})^2$. Then we have
\begin{equation}
    f_{\kappa}(x)=\exp{(\frac{-\kappa}{a(x+\frac{b}{2a})})}.
\end{equation}
Note that $f_{\kappa}$ could be multivalued due to the logarithm if $\Delta\ne 0$. At the moment, we keep this ambiguity and determine how to deal with it later.

Once we know the $f_{\kappa}$'s, from the expression (\ref{eqn:ellkappadef}), it is straightforward to calculate the following commutation relation:
\begin{equation}
[\ell_{\kappa}^{(+)}, \ell_{\kappa'}^{(+)}]=(\kappa-\kappa')\ell_{\kappa+\kappa'}^{(+)},
\end{equation}
establishing the representation of the Witt algebra or the classical Virasoro algebra in the form of the differential operators on $x^{+}$.

We can proceed and define the Virasoro generators as
\begin{equation}
{\cal L}^{(+)}_{\kappa}\equiv  N
\int dx^{+} g(x^{+})f_{\kappa}(x^{+})T_{++}(x^{+}), \label{eqn:calLdef}
\end{equation}
where $T_{++}$ is $++$ component of the energy momentum tensor, and $N$ is the normalization constant to be fixed later.

\noindent {\bf Virasoro Algebra} Postulating Wightman axioms, which include Poincar\'e symmetry, and dilatation symmetry, imposes several restrictions on the energy momentum tensor; $T_{+-}=T_{{-+}}=0$, and $T_{\pm\pm}$ depends only $x^{\pm}$ . This fact was used in the expression in Eq. (\ref{eqn:calLdef}). It can be further derived that the commutation relations among the components of the energy momentum tensor takes the following form \cite{Luscher:1976}:
\begin{eqnarray}
[T_{++}(x^{+}),T_{++}(y^{+})]&=&\frac{\cc}{6\pi}i^{3}\delta'''(x^{+}-y^{+})+4i\delta'(x^{+}-y^{+})T_{++}(y^{+}) \nonumber \\
& &\left. -2i\delta(x^{+}-y^{+})\partial_{y}T_{++}(y)\right|_{y=y^{+}} \, , \label{eqn:TTcom}
\end{eqnarray}
where $\cc\geq 0$ is a constant that would characterizes the conformal field theory in question. The same commutation relation can be obtained for $T_{--}$. 
The commutation relations among the Virasoro generators  follow from Eq. (\ref{eqn:TTcom}) directly:
\begin{equation}
\left[{\cal L}^{(+)}_{\kappa}, {\cal L}^{(+)}_{\kappa'}\right]=2iN(\kappa'-\kappa) {\cal L}^{(+)}_{\kappa+\kappa'} +\frac{\cc iN^{2}}{6\pi}{\cal I}[\kappa|\kappa'],
\end{equation}
where
\begin{equation}
    {\cal I}[\kappa|\kappa'] \equiv \int dx \left\{\kappa^{3}+\left(2gg''-(g')^{2}\right)\kappa + g^{2}g''' \right\} \frac{f_{\kappa+\kappa'}}{g} . \label{eqn:Ikk'def}
\end{equation}
The above expression ${\cal I}[\kappa|\kappa']$ can be further simplified when $g$ is a quadratic function (\ref{eqn:gdef}) as follows:
\begin{eqnarray}
 {\cal I}[\kappa|\kappa']&=& (\kappa^3-\Delta\kappa)\int \frac{dx}{g(x)}\exp\left[{\left(\kappa+\kappa'\right)\int^x\frac{dx'}{g(x')}}\right]  \\ 
 &=&  (\kappa^3-\Delta\kappa)\int^{\chi_f}_{\chi_i} d\chi e^{(\kappa+\kappa')\chi} ,\label{eqn:Ikk'bychi}
\end{eqnarray}
if we introduce a new variable $\chi$ as
\begin{equation}
    \chi\equiv \int^x\frac{dx'}{g(x')},\label{eqn:chidef}
\end{equation}
and specify the interval of the integration.
We may further breakdown Eq. (\ref{eqn:Ikk'bychi}) as follows:
\begin{equation}
 {\cal I}[\kappa|\kappa']=\left\{
 \begin{array}{ll}
\left.  \frac{\kappa^3-\Delta\kappa}{\kappa+\kappa'}e^{(\kappa+\kappa')\chi}\right|^{\chi_f}_{\chi_i}
  &  \hbox{for} \ \ \kappa + \kappa'\neq 0 \\ \ & \ \\
  \left. (\kappa^3-\Delta\kappa) \chi \ \right|^{\chi_f}_{\chi_i}
  & \hbox{for}\ \  \kappa+\kappa'=0    \end{array}\right. . 
  \label{eqn:Ibychi}
\end{equation}
Since $\chi_f$ and $\chi_i$ would depend on the roots of $g$, in particular its discriminant $\Delta$, also minding the relation between $\Delta$ and the Casimir invariant, we analyze the following three cases separately depending on the signature of $\Delta$.

\noindent$\mathbf{\Delta<0}$ First, we are concerned with the case $\Delta < 0$.
In this case, two roots $x_{\langle \pm \rangle}$ are complex numbers. If we write
\begin{equation}
    x- x_{\langle \pm \rangle}\equiv R_{\pm}e^{i\theta_\pm},
\end{equation}
from Eq. (\ref{eqn:chidef}),
\begin{equation}
    \chi=\frac{i}{\sqrt{\Delta}}\left(\theta_+ - \theta_- \right).
\end{equation}
It is easy to see that  $\chi$ covers $0$ to $\frac{2\pi i}{\sqrt{\Delta}}$ on the real axis, as $x$ varies from $-\infty$ to $\infty$. However, there is apparently ambiguity of $\frac{2\pi ni}{\sqrt{\Delta}} (n\in \mathbb{Z})$, that stems from the choice of the principal value of $\theta_\pm$. This ambiguity reflects on the component $f_\kappa$ of the differential operators we introduced in Eq (\ref{eqn:ellkappadef}) as
\begin{equation}
    f_\kappa=e^{\kappa\chi}.
\end{equation}
To circumvent the predicament, we restrict the index of the differential operators $\kappa$ as follows:
\begin{equation}
    \kappa = \sqrt{\Delta}m \ , \ \ m\in \mathbb{Z}.
\end{equation}
With this restriction, the ambiguity in the differential operators $\ell_\kappa^{(+)}$ is resolved and we obtain
\begin{equation}
 {\cal I}[\kappa|\kappa']=2\pi i \Delta (m^3-m)\delta_{m+m',0}.
\end{equation}

Therefore, relabelling the Virasoro generators as $L_m^{(+)}\equiv{\cal L}_\kappa^{(+)}$ and choosing the normalization $N=-\frac{1}{2i\sqrt{\Delta}}$, we arrive at the familiar expression of the Virasoro algebra:
\begin{equation}
    \left[{L}^{(+)}_{m}, { L}^{(+)}_{m'}\right]=(m-m') {L}^{(+)}_{m+m'} +\frac{\cc }{12}(m^3-m)\delta_{m+m',0}\ .
\end{equation}
This is the reproduction of the result by L\"uscher and Mack \cite{Luscher:1976} 
\footnote{To match the expression exactly as in the note by L\"uscher \cite{Luscher:1976}, we need to adopt $\kappa=-\sqrt{\Delta}m$.}, since $\Delta<0$ case includes L\"uscher-Mack Hamiltonian (\ref{eqn:c2-1}). In particular, the above shown Virasoro algebra does not depend on the details of the construction such as particular coefficients of the function $g$ except the signature of $\Delta$.

\noindent$\mathbf{\Delta>0}$ Next we move to the case $\Delta>0$. In this case, two poles $x_{\langle\pm\rangle}$ reside on the real axis in regard to $g$. In order to avoid the ambiguity in the integral, it would be natural to restrict the interval of the integration between the two poles. Still, it is apparent that there would be divergence at the poles, from Eqs. (\ref{eqn:chidef}) and (\ref{eqn:Ikk'def}). Therefore we additionally introduce a cut-off $\epsilon \ll 1$ near $x_{\langle\pm\rangle}$ in the interval of the integration:
\begin{equation}
    [x_{\langle-\rangle}+\epsilon,x_{\langle+\rangle}-\epsilon].
\end{equation}
The treatment in the Euclidean case \cite{Tada:2019rls} also inspires the introduction of the cutoff.

From Eq. (\ref{eqn:Ibychi}), ${\cal I}[\kappa|\kappa']$ for $\kappa+\kappa'\neq 0 $ can be estimated as follows:
\begin{equation}
  {\cal I}[\kappa|\kappa']_{\kappa\neq-\kappa'}=\frac{\kappa^3-\Delta\kappa}{\kappa+\kappa'}
  \left(e^{\frac{\kappa+\kappa'}{\sqrt{\Delta}}\left(\ln (\epsilon a / \sqrt{\Delta})+i\pi\right)}
  -e^{\frac{\kappa+\kappa'}{\sqrt{\Delta}}\left(-\ln (\epsilon a /\sqrt{\Delta})+i\pi\right)} \right),\label{eqn:Ikk'cond}
\end{equation}
which should be null in order for the Jacobi identity of the Virasoro algebra to be held. This requirement can be met if we demand that the index $\kappa$ satisfies the following condition\footnote{We could actually allow $\kappa$ to be multiple of a half-integer to annihilate Eq. (\ref{eqn:Ikk'cond}), but then, we would not have ${\cal L}_0$ which is supposed to be the Hamiltonian.}:
\begin{equation}
    \kappa = \frac{\pi i\sqrt{\Delta}}{\ln{\frac{\epsilon a}{\sqrt{\Delta}}}}n, \:\:\:\: n\in\mathbb{Z } \:\:\:
\label{eqn:kappan}
\end{equation}
Note that the above condition conveniently resolves the problem of multivaluedness of $f_\kappa$, which we encountered in the $\Delta <0$ case. 

When $\kappa+\kappa'=0$, 
\begin{eqnarray}
   {\cal I}[\kappa|-\kappa] &=&   \frac{\kappa^3-\Delta\kappa}{ \sqrt{\Delta}}2\ln{\frac{\epsilon a}{\sqrt{\Delta}}}\\
   &=&\frac{-2i\pi^3\Delta}{\left(\ln{\frac{\epsilon a}{\sqrt{\Delta}}}\right)^2}\left(n^3+\frac{\left(\ln{\frac{\epsilon a}{\sqrt{\Delta}}}\right)^2}{\pi^2}n\right).
\end{eqnarray}
Choosing the following somewhat intricate expression as normalization $N$:
\begin{equation}
    N= \frac{\ln{\frac{\epsilon a}{\sqrt{\Delta}}}}{2\sqrt{\Delta}\pi},
\end{equation}
we arrive at the commutation relation
\begin{equation}
  \left[L_n^{(+)},L_{n'}^{(+)}\right]= (n-n'){L}^{(+)}_{n+n'}+
    \frac{\cc}{12}\left( n^3+\frac{\left(\ln{\frac{\epsilon a}{\sqrt{\Delta}}}\right)^2}{\pi^2}n\right)\delta_{n+n',0}\ \ ,
\end{equation} 
if we introduce the notation $L^{(+)}_n={\cal L}^{(+)}_\kappa$.
Note that there is a divergence in the central charge term. This divergence, however, can be absorbed by shifting $L_0^{(+)}$ as follows:
\begin{equation}
    {L}^{(+)}_{0}\to L^{(+)}_{0}+\frac{\cc}{24}\frac{\left(\ln{\frac{\epsilon a}{\sqrt{\Delta}}}\right)^2}{\pi^2}
\end{equation} This shift could be related to the Schwarzian derivative term mapping to a torus in Euclidean setup.
All in all, we have the following Virasoro algebra for $\Delta>0$:
\begin{equation}
  \left[L_n^{(+)},L_{n'}^{(+)}\right]= (n-n'){L}^{(+)}_{n+n'}+
    \frac{\cc}{12} n^3\delta_{n+n',0}\ \ \ n,n' \in \mathbb{Z}\
    . \label{eqn:VirasoroDeltaPlus}
\end{equation}

We would obtain the Rindler Hamiltonian $\hat{M}_{01}$ (\ref{eqn:c21}) by choosing $b=1$, $c=0$ and taking $a\to0$. Which means the distance between the two poles is infinite. The Virasoro generators can be obtained by applying suitable rescaling of $\kappa$ and choice of $N$ in the following expression:
\begin{equation}
    {\cal L}^{(+)}_{\kappa}\equiv N\int dx^+ x^+(x^+)^{\kappa} T_{++}(x^+).
\end{equation}
Note that in the above $\kappa$ is actually a imaginary number as one can see from Eq. (\ref{eqn:kappan}).


\noindent$\mathbf{\Delta=0}$ Lastly, we turn our attention to the most intriguing $\Delta=0$.  When $\Delta=0$, we have $g(x^+)=a (x^+-x_r)^2$ where $x_r=-\frac{b}{2a}$. Thus we have \begin{equation}
    \chi=\int^{x^+} \frac{dx}{g(x)}=-\frac{1}{a(x^+-x_r)}.
\end{equation}
It is apparent that as $x^+$ runs from $-\infty$ to $\infty$, $\chi$ also takes all the values between $-\infty$ and $\infty$, thus 
\begin{equation}
    \chi_i=-\infty \ \ , \ \ \ \chi_f=\infty.
\end{equation}

It follows from Eq. (\ref{eqn:Ikk'bychi}) that
\begin{equation}
    {\cal I}[\kappa|\kappa']=\kappa^3 \int^\infty_{-\infty} d\chi e^{(\kappa+\kappa')\chi}.
\end{equation}
It is apparently natural for us to write $\kappa=i k$ where $k\in \mathbb{R}$ and derive the delta function $2\pi\delta(k+k')$. Redefining ${\cal L}^{(+)}_{k}\equiv{\cal L}^{(+)}_{\kappa}$ and choosing $N=\frac{1}{2}$, we would arrive at
\begin{equation}
    [{\cal L}^{(+)}_{k},{\cal L}^{(+)}_{k'}]=(k-k'){\cal L}^{(+)}_{k+k'}+\frac{ck^3}{12}\delta(k+k'), \:\:\: k\in \mathbb{R}
\end{equation}
We see that in the degenerate $\Delta=0$ case, the Virasoro algebra would evoke the continuum spectrum, because the index of the Virasoro algebra takes continuous real value just as found in the study of the sine-square deformation of the Euclidean conformal field theories \cite{Ishibashi:2015jba,Ishibashi:2016bey}.

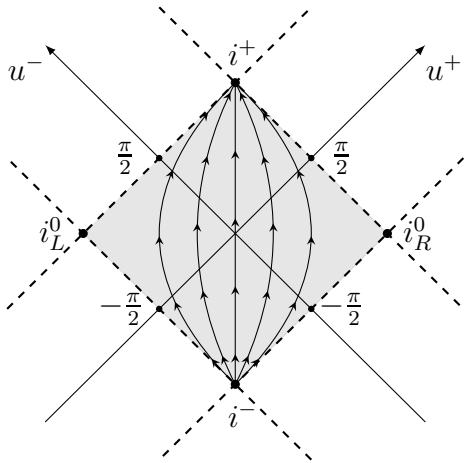
\begin{figure}[tbh]
\centering
\begin{tikzpicture}[>=latex,decoration={
markings,mark=between positions 0.1 and 0.97 step 0.9cm with {\arrow{stealth}}}]
\draw[->](-2.5,-2.5)--(2.5,2.5);
\draw[->](2.5,-2.5)--(-2.5,2.5);

\draw[thick,dashed](-1,3)--(3,-1);
\draw[thick,dashed](3,1)--(-1,-3);
\draw[thick,dashed](1,-3)--(-3,1);
\draw[thick,dashed](-3,-1)--(1,3);

\draw(2.75,2.2) node {$u^+$};
\draw(-2.75,2.2) node {$u^-$};
\draw(1.4,1) node {$\frac{\pi}{2}$};
\filldraw (1,1) circle (1pt);
\draw(-1.5,-1) node {$-\frac{\pi}{2}$};
\filldraw (-1,-1) circle (1pt);
\draw(1.4,-1) node {$-\frac{\pi}{2}$};
\filldraw (1,-1) circle (1pt);
\draw(-1.45,1) node {$\frac{\pi}{2}$};
\filldraw (-1,1) circle (1pt);
\draw(0.1,2.4) node {$i^+$};
\filldraw (0,2) circle (1.5pt);
\draw(0.1,-2.4) node {$i^-$};
\filldraw (0,-2) circle (1.5pt);
\draw(-2.4,0) node {$i_L^0$};
\filldraw (-2,0) circle (1.5pt);
\draw(2.4,0) node {$i_R^0$};
\filldraw (2,0) circle (1.5pt);
\filldraw[opacity=0.1] (0,2)--(2,0)--(0,-2)--(-2,0);
\draw [postaction={decorate}] (0,-2)--(0,2);

\draw [postaction={decorate}] [rotate around={90:(0,0)}] (-2,0) parabola bend (0,1) (2,0);
\draw [postaction={decorate}] [rotate around={90:(0,0)}] (-2,0) parabola bend (0,0.5) (2,0);
\draw [postaction={decorate}] [rotate around={90:(0,0)}] (-2,0) parabola bend (0,-1) (2,0);
\draw [postaction={decorate}] [rotate around={90:(0,0)}] (-2,0) parabola bend (0,-0.5) (2,0);

\end{tikzpicture}
\caption{The universal covering space of the Penrose diamond (shaded region) is charted by $u^+$ and $u^-$. While the conformal symmetry is represented utilizing the entire covering space, the time flow by $P_0$ (arrowed line) is confined within a single Penrose diamond.}
\label{fig:D=0Penrose}
\end{figure}

The physical implication of the continuous Virasoro algebra and the consequential continuous spectrum is described in the following. Since the above  analysis applies to the case that we choose $\hat{P}_0$ as the Hamiltonian, by taking the limit $a,b\to 0$, we explain using  $\hat{P}_0$ in Eq. (\ref{eqn:defP0}) as the Hamiltonian. The continuum spectrum indicates that the system contains at least either infinite space or infinite time. Although Minkowski spacetime is infinite both in space and time, the covering space structure required by the conformal symmetry imposes effectively compact spacetime, which can be best elucidated in terms of $u^\pm$ coordinates as in Eq. (\ref{eqn:xpmupm}). This is the reason why L\"uscher-Mack Hamiltonian shows the discrete spectrum as implied by the discrete Virasoro algebra. On the other hand, the flow of time induced by $\hat{P}_0$ becomes zero at $u^\pm=\pm\frac{\pi}{2}$ as one can readily see from Eq. (\ref{eqn:defP0}). Hence time does not flow beyond a single Penrose diamond (Fig. \ref{fig:D=0Penrose}). Despite the existence of covering space, which is imperative to accomodate the conformal symmetry, the system quantized by the Hamiltonian with $\Delta=0$ can not see the non-simply connected structure of spacetime. As a result, the system virtually retrieves infinite spacetime.

In the study of the SSD of Euclidean conformal field theory \cite{Ishibashi:2015jba,Ishibashi:2016bey}, the corresponding quantization was called dipolar quantization as opposed to usual radial quantization; The emergence of the continuum Virasoro algebra and infinite space was observed. Obviously, the same physical mechanism is working in the Lorentzian case.

In summary, we have analyzed three types of Hamiltonians of 2d Lorentzian CFT and derived the respective Virasoro algebra. We have reproduced the result by L\"uscher and Mack \cite{Luscher:1976} for the negative Casimir invariant. The positive Casimir invariant is related to the Rindler Hamiltonian, and we also derived the discrete Virasoro algebra for this case. 

Most interestingly, for the zero Casimir invariant, we obtained the continuous Virasoro algebra and infinite spacetime. This result would clarify some of the confusion regarding Lorentzian conformal field theories. Since the conformal invariance requires the introduction of the universal covering space, sometimes the Lorentzian conformal field theory is considered unphysical due to the existence of a closed time-like curve. Our analysis makes it clear that there is no closed time-like curve if we choose naive time-translation $\hat P_0$ instead of L\"uscher-Mack Hamiltonian.


\vspace{1cm}
\noindent{\bf Acknowledgment:} The authors would like to thank E. Itou, H. Katsura, H. Kawai, N. Ishibashi, Y. Kubota, M. Nozaki, K. Okunishi, S. Ryu, T. Takahashi, 
X. Wen, G. Wong, 
 and the participants of the iTHEMS workshop ``Workshop on Sine square deformation and related topics,'' for fruitful discussions,  comments, and suggestions, which significantly contributed to the present work. 
XL would like to thank H. Liang for his guidance and encouragement.
TT is in part supported by JSPS Kakenhi Grant Number JP19K03679.

\end{document}